\documentclass[12pt]{article}
\usepackage{graphicx}

\newcommand\pubnumber{}
\newcommand\pubdate{3 July, 2015}

\newcommand{\beq}{\begin{equation}}
\newcommand{\eeq}{\end{equation}}
\newcommand{\beqa}{\begin{eqnarray}}
\newcommand{\eeqa}{\end{eqnarray}}
\newcommand{\bcent}{\begin{center}}
\newcommand{\ecent}{\end{center}}
\newcommand{\hiroshima}{Graduate School of Science, Hiroshima University, Kagamiyama, Higashi-Hiroshima 739-8526, Japan}
\newcommand{\izest}{International Center for Zetta-Exawatt Science and Technology, Ecole Polytechnique, Route de Saclay, Palaiseau, F-91128, France}
\newcommand{\icr}{Institute for Chemical Research, Kyoto University Uji, Kyoto 611-0011, Japan}

\def\Title#1{\begin{center} {\Large #1 } \end{center}}
\def\Author#1{\begin{center}{ \sc #1} \end{center}}
\def\Address#1{\begin{center}{ \it #1} \end{center}}

\newcommand\pubblock{\rightline{\begin{tabular}{l} \pubnumber\\
         \pubdate  \end{tabular}}}
\newenvironment{Abstract}{\begin{quotation}  }{\end{quotation}}
\newenvironment{Presented}{\begin{quotation} \begin{center} 
             PRESENTED AT\end{center}\bigskip 
      \begin{center}\begin{large}}{\end{large}\end{center} \end{quotation}}

\def\beq{\begin{equation}}
\def\eeq#1{\label{#1}\end{equation}}
\def\eeqn{\end{equation}}

\def\beqa{\begin{eqnarray}}
\def\eeqa#1{\label{#1}\end{eqnarray}}
\def\eeqan{\end{eqnarray}}

\let\bar=\overbar

\def\Dslash{\not{\hbox{\kern-4pt $D$}}}
\def\dslash{\not{\hbox{\kern-2pt $\del$}}}

\def\msb{{\bar{\ssstyle M \kern -1pt S}}}

\begin{document}
\begin{titlepage}
\pubblock

\vfill
\Title{Gamma Polari-Calorimetry with SOI pixels for proposals at Extreme Light Infrastructure (ELI-NP)}
\vfill
\Author{Kensuke Homma${}^{1,2}$ and Yoshihide Nakamiya${}^3$\\ on behalf of RA5 at ELI-NP}
\Address{${}^1$\hiroshima \\ ${}^2$\izest \\ ${}^3$\icr}
\vfill
\begin{Abstract}
We introduce the concept of Gamma Polari-Calorimetry (GPC) dedicated for proposals 
at Extreme Light Infrastructure in the Romanian site (ELI-NP). A simulation study shows
that an assembly of thin SOI pixel sensors can satisfy our requirements to GPC.
\end{Abstract}
\vfill
\begin{Presented}
International Workshop on SOI Pixel Detector (SOIPIX2015),
Tohoku University, Sendai, Japan,\\ 3-6, June, 2015.
\end{Presented}
\vfill
\end{titlepage}
\def\thefootnote{\fnsymbol{footnote}}
\setcounter{footnote}{0}

\section{Introduction}
From a field theoretical point of view, 
high energy density states created in the vacuum are interesting 
objects to be studied.
If the energy density is high enough, particles and anti-particles are created
by converting the energy into particle masses and momenta.
In order to understand such particle production mechanisms from intense fields, the
perturbative treatment of elemental interactions by expanding with weak coupling strengths 
is inappropriate, instead, the nonperturbative approach by expanding with field strengths
is rather important. 
So far such high energy density states in the vacuum
are realized only in point-like or quasi-point-like 
spots by colliding high energy charged particles where
produced systems are maintained only for an instant and rapidly evolve into 
dynamically complicated final states.
This makes comparisons between theoretical predictions and experimental data difficult.
In this respect, elemental dynamcs in strong fields in the vacuum
are not quantitatively understood yet. High-intensity laser fields, on the other hand, 
can produce relatively static high energy density states with controlled polarization states
where spontaneous $e^+e^-$ pair creations are expected.
Moreover, a high field laser also simply implies that we can use 
a large number of laser photons or high luminosity of photon beams. 
This allows us to search for undiscovered low-mass fields weakly coupling to photons
which could be relevant to dark components of the Universe~\cite{DEptp,DEptep,hiroshima,kyoto1}.

The pair creation threshold is known as the Schwinger critical field~\cite{Schwinger}
$E_s = \frac{m^2 c^3}{e\hbar}=1.3 \times 10^{18} \mbox{V/m}$
corresponding laser intensity of $4\times 10^{29} \mbox{W/cm}{}^2$.
An electric field $E$ causes the pair creation via the tunneling effect in the vacuum.
The rate of the pair creation in the electric field $E$ is proportional to the following
tunneling factor $\exp\{-\pi E_s/E\}$. This corresponds to the non-perturbative 
description of the phenomenon.

\begin{figure}
\bcent
\includegraphics[width=9.0cm]{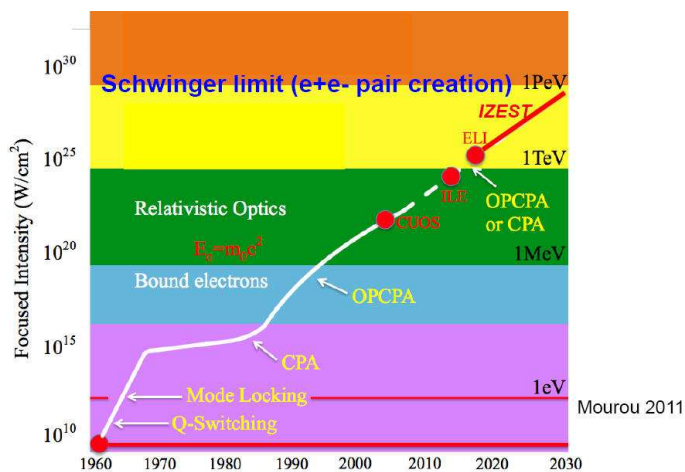}
\caption{
The leap of laser intensity as a function of year\cite{TM}.
}
\label{Fig1}
\ecent
\end{figure}

The leap of high intensity lasers as a function of year is illustrated in Fig.1.
As one of cutting-edge projects, Extreme Light Infrastructure (ELI)~\cite{ELI}
is approved by European Union which has been eventually divided into three sites: 
ELI-Beamlines facility in Czech Republic, ELI-Attosecond facility in Hungary and 
ELI-Nuclear Physics facility (ELI-NP) in Romania.
In particular, the ELI-NP site is a unique laser facility where two 10 PW (220 J / 22fs) lasers 
operated every 60 sec and also a 700 MeV electron beam from a linac are simultaneously 
available within the same experimental area~\cite{ELI-NP}.
With a 10 PW field, we may be able to achieve $10^{22-24}$ W/cm${}^2$ depending on the 
achievable spot size of the focused beam. This is still far from $E_s$. 
Because of the strong exponential suppression, we do not expect
for tunneling pair creations to occur with a reasonable event rate in the laboratory.
However, if high-intensity lasers could be combined with energetic photons or electrons, 
we can effectively lower the tunneling probability for the pair creation 
from the laser induced vacuum. The production rate $R_{e^+e^-}$ is evaluated as
$
R_{e^+e^-} = \frac{e^2 {E_l}^2}{4\pi^3}
\exp \{
-\frac{8}{3} \frac{E_{s}}{E_l} \frac{m_e c^2}{\hbar\omega_{\gamma}}
\}
$
where $E_l$ is the electric field of laser and $\hbar\omega_{\gamma}$ is the
incident $\gamma$-ray energy~\cite{Narozhny}.

How can we prepare the high energy electron probe first ? 
We expect that a high-intensity laser can be used to generate electrons
based on the Laser-Plasma-Accelerator technology. Recently 4.2 GeV
electron acceleration is demonstrated with 0.3 PW ~\cite{LPA} 
and actually it is planned to be extended beyond 10~GeV at the ELI-Beamlines facility. 
If we can use a multi-GeV electron beam, we also can produce high energy $\gamma$-rays via
Compton scattering between high energy electrons and a high-intensity laser field.
Especially in the nonlinear Compton regime, more energetic $\gamma$-rays can be produced
due to multi-photon absorption processes compared to that of the linear Compton scattering.
However, in 10 PW class lasers, the nonlinear Compton process itself is subject to be verified.
Therefore, the measurement of the degree of linear polarization as well as scattered photon and
electron energies will be important by itself and also for later applications as a secondary
$\gamma$-ray source.

Even before reaching the real pair creation threshold, 
we can discuss the nonperturbative aspect of interactions 
between a probe photon and a high-intensity field by measuring 
so-called vacuum birefringence where refractive index of the vacuum under a high-intensity laser field 
depends on the polarization direction of probe photons~\cite{APB-QED}. 
With the ELI-NP parameter, the achievable refractive index change from that of the
vacuum is only on order of $10^{-9}$, therefore, 
it is extremely difficult to probe it with the optical wave length.
Although such difficulty might be compensated by high statistics of probe laser photons per shot
as discussed in~\cite{APB-QED},
as an alternative probe, we may consider to utilize polarized $\gamma$-rays to enhance
the phase shift with a much shorter wave length.
For this purpose, we have to be able to measure the degree of linear polarization of
incident probe $\gamma$-rays.

Given these purposes and environments at ELI, 
it is indispensable to develop a detection system which
allows us to measure the degree of linear polarization of produced $\gamma$-rays as well as their
energies simultaneously. This is the motivation for us to develop Gamma Polari-Calorimeter (GPC).

\begin{figure}
\bcent
\includegraphics[width=9.0cm]{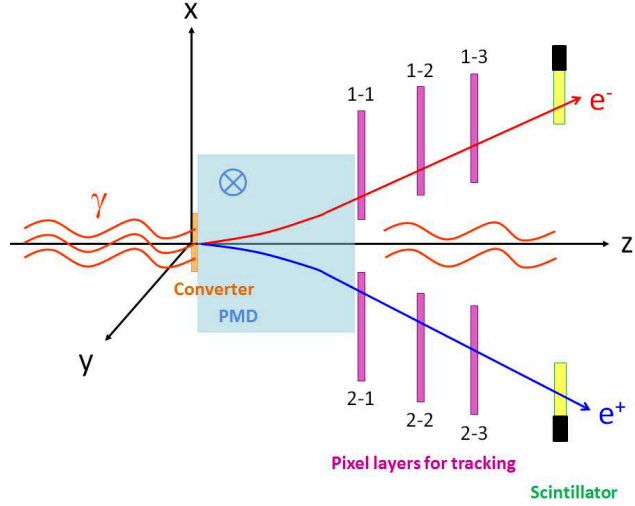}
\caption{
Schematic view of Gamma Polari-Calorimeter (GPC).
}
\label{Fig2}
\ecent
\end{figure}

\section{Gamma Polari-Calorimeter (GPC)}
Requirements to  GPC are 
\begin{description}
\item[1)] measurability of polarization from 0.1 GeV to a few GeV,
\item[2)] charge separations between electrons and positrons with momentum resolution below 20~MeV,
\item[3)] resolvable multiple $\gamma$-ray incidence up to order of ten $\gamma$-rays,
\item[4)] compact enough to make the detection system movable,
\item[5)] readout speed of 10 Hz at most.
\end{description}
Special conditions at ELI-NP are 
\begin{description}
\item[I)] $\gamma$-rays are generated from a laser focal spot within a cone angle of sub-mrad,
\item[II)] order of $10^4$ $\gamma$-rays around 1 GeV are injected per shot.
\end{description}

Due to the energy range required in 1),
we can utilize the photon conversion process within a thin foil in front of
a uniform dipole magnetic field as illustrated in Fig.2.
Since $e^+e^-$ pairs are created within the foil, 
we can measure the momenta of the created pairs, equivalently, energies of incident $\gamma$-rays
and also anisotropy of angles of reaction planes including generated pairs
with respect to the linear polarization plane of incident $\gamma$-rays.
Thus requirements 1) and 2) are satisfied with the conversion-based design.
Because multiple scatterings within the foil dominantly degrades the angular resolution
as well as momentum resolution, the thickness of the foil should be kept
as thin as possible. On the other hand, the thickness also changes the conversion rate. 
By assuming II), the thickness of high-Z foil is most likely to be 20~$\mu$m with Au atoms
in order to satisfy the requirement 3).
Due to the special condition I), the conversion vertex point is well
localized within a typical pixel size. Therefore, the accuracy of the track reconstruction
is essentially determined by the pixel resolution of the first layer sensor
and uniformity of the dipole magnetic field.
We plan to introduce a compact Halbach-type parmanent-magent-based dipole
in order to satisfy the requirement 4). 
The thickness of a pixel sensor layer is also important to avoid further increase
of the number of multiple scatterings in order to allow accurate extrapolations
of straight tracks to the first layer hit points.
A square shaped sensor element is preferable compared to a rectangular shaped element
like a strip-type sensor
for the measurement of anisotropy of the reaction plane, because it makes analysis
simpler without increasing the total thickness.
Finally the requirement 5) is achievable if we assume to use a set of INTPX4 SOI sensor chips
and the readout system~\cite{INTPIX4}.

\section{Conclusion}
We performed Geant4-based simulation by assuming available INTPIX4 SOI sensor 
chip with the pixel size of $17 \times 17$ $\mu$m${}^2$ with geometry in Fig.3.
The obtained momentum resolution to electron or positron tracks above 0.2 GeV
is at a 7\% level and the analyzing power of the polarimeter is about 50\%. 
We thus conclude that the INTPIX4-based sensor assembly
can surely perform polarimetry and calorimetry with required resolutions at 
the ELI-NP proposals.

\begin{figure}
\bcent
\includegraphics[width=9.0cm]{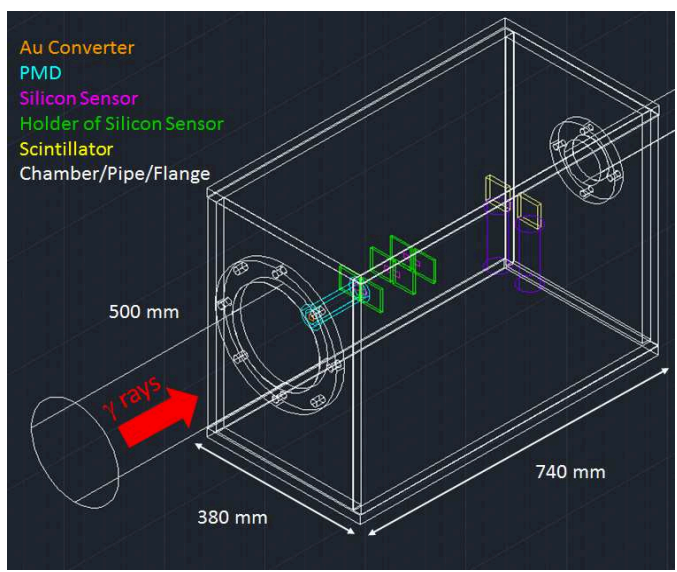}
\caption{
A present design of Gamma Polari-Calorimeter (GPC).
}
\label{Fig3}
\ecent
\end{figure}

\end{document}